\documentclass[11pt,twoside]{article}

\usepackage{asp2006}

\usepackage{fancyhdr,graphicx,caption,rotating,multirow,lscape}
\DeclareGraphicsExtensions{.eps,.eps.gz,.ps,.jpg,.tiff}

\begin{document}

\title{Precision photometry for planetary transits}  

\author{F. Pont}  
\affil{Geneva University Observatory}    
\author{C. Moutou}
\affil{Laboratoire d'Astrophysique de Marseille}

\begin{abstract} 
This paper reviews the state of the art in follow-up photometry for planetary transit searches. Three topics are discussed: (1) Photometric monitoring of planets discovered by radial velocity to detect possible transits (2) Follow-up photometry of candidates from photometric transit searches to weed out eclipsing binaries and false positives (3) High-precision lightcurves of known transiting planets to increase the accuracy on the planet parameters.
\vspace{2cm}
\end{abstract}


\section{Introduction}

Photometric surveys for transiting planets have been pushing the frontier of wide-field, high-cadence photometry at the millimagnitude level, as illustrated by several contributions to this volume. A different mode of photometry --lightcurve measurement for a single object at the expected time of the transit -- is also essential to the field of transiting planets, for three specific purposes:
\smallskip \smallskip

\noindent
1- to determine whether a   planet discovered by radial-velocity surveys is transiting or not
 \smallskip \smallskip

\noindent
2- to sort out the transit candidates produced by a photometric search, weeding out eclipsing binaries and other false positives.
\smallskip \smallskip

\noindent
3- to obtain a precise determination of the radius and density of  known transiting planets.\smallskip \smallskip

We discuss here the performance and limitations of these three modes, together with their current applications and future perspectives.

\newpage

\section{Search for transits on radial-velocity planets}

Given the importance of planetary transits to the understanding of extrasolar planets, all short-period planets found by radial velocity surveys are measured in photometry at the expected time of alignment of the star and planet with the line-of-sight. 

Until recently, the planets detected by radial velocity surveys had masses comparable to Saturn or larger, and were therefore thought to be gas giants larger than 60,000 km in radius.  Such planet transiting across the disc of a Solar-type star would produce a transit signal near the 1\% level. 
For bright stars (radial-velocity search targets usually have $V<8$), finding or excluding a transit signal at the 1\% level is within the capacity of small telescopes from the ground.  The first planetary transit outside the Solar System was found in such a way (Charbonneau et al. 2000), with a 10-cm camera on a ``backyard'' telescope  on the planet identified in radial velocity by Mazeh et al. (2000). Typically, the detection limit of photometric checks for transits on radial velocity planets is at the level of a few millimagnitudes.


\begin{figure}[!th]
\centering
\includegraphics[width=7cm]{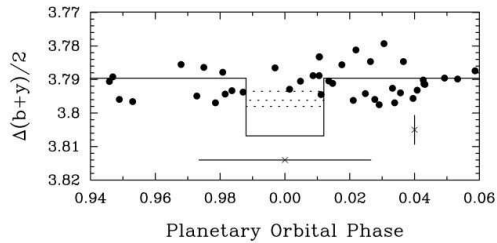}
\caption{Example of lightcurve at the predicted epoch of line-of-sight star-planet alignment  for a planet detected by radial velocity (reproduced from Butler et al. 2004). The lines sketch the predicted transits for different planet density. The horizontal bar gives the uncertainty on the transit timing.}
\end{figure}

Two other transiting planets were identified this way: HD149026b (Sato et al. 2005) and HD 189733b (Bouchy et al. 2005). The objects are of considerable importance among the known transiting exoplanets, because they orbit bright stars ($V\sim$ 7), and therefore many derivative studies are possible.
HD149026b brought with it a surprising realisation: although its mass is comparable to that of Saturn, it is markedly smaller in size. The reason is thought to be the presence of a much larger amount of heavy elements in its composition.  This planet produces a transit of only 0.3\% in the lightcurve of its host stars (0.72 R$_J$ in front of a G0IV star), which begs the obvious question:  \\ \smallskip
{\it could we have missed any transiting gas giant among the known hot Jupiters?}

The answer is yes for several reasons: large-core, small hot Jupiter could have been missed by producing shallower transits than expected; planets can also transit their star at high latitude, which produces a dimming that is both smaller and shorter, rendering the detection more difficult; finally, sometimes the phasing of the transit has an uncertainty of several hours, due to the radial velocity solution, so that the transit may have been missed because it did not take place near enough to the predicted time.

\begin{figure}[!th]
\centering
\includegraphics[width=9cm]{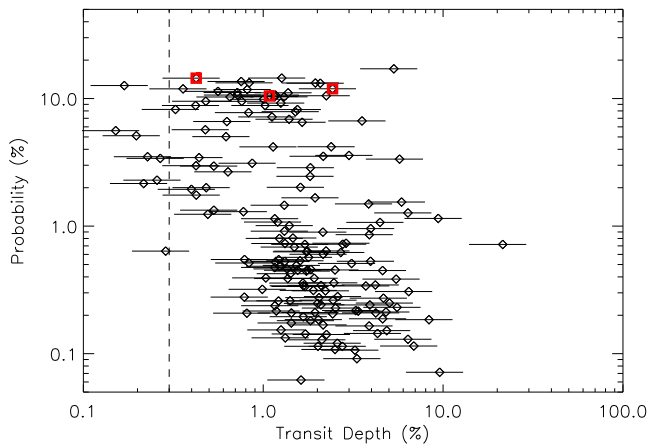}
\caption{Transit probability versus depth for planets detected by
  radial-velocity surveys. The size of horizontal bars shows the depths for a
  range of planetary densities (0.5 to 1.5 g$\cdot$cm$^{-3}$). Square symbols indicate transiting planets. The dashed line is the approximate limit of ground-based photometry.}
\label{tprob}
\end{figure}

However, statistically, there is no compelling indication that any transiting hot Jupiters was missed in the present radial-velocity sample. At the time of detection of HD209458$b$, there were 25 known extrasolar planets, 8 of which were in orbit shorter than 10 days, with up to 10\% probability to have the correct alignement for transit detection. The cumulative probability of transit was therefore near unity. Today, the three transiting planets identified from radial-velocity surveys stand among 53 known planets with a semi-major axis smaller than 0.1 AU. Again, the cumulative transit probability is near 3. Nevertheless, it may be worth taking a second look at the issue.

In the past couple of years, radial velocity instruments have reached the meter-per-second accuracy, which has opened the way to the detection of exoplanets of smaller and smaller mass - a dozen objects are now known with masses in the Neptune/Uranus range and below. Transit search follow-up photometry  on ``hot Neptunes'' found by RV surveys is therefore a serious contender in the hunt for the first telluric planet transits, together with the space transit searches Corot and Kepler.

However, composition models show that these objects could have sizes in the range of 2-3 Earth diameters. Some of them orbit an M dwarf, and in this case the transit could be deep enough for detection from the ground (see e.g. Lopez-Morales et al. 2006  for GJ581b). But if the host star is solar, a transiting ``hot Neptune'' can produce dimmings of 0.1\% or shallower. Precise differential photometry of stars brighter than 8th magnitude from the ground requires a relatively small telescope to avoid saturation and a field-of-view of half a degree or more to get suitable comparison stars. In these conditions, it is very difficult to reach the 0.1\% level.  Even if the necessary photon signal can be gathered, correlated noise due to the atmosphere will mask any signal of this amplitude.  One solution is to observe these objects from space, which such facilities as the MOST Canadian satellite or the Huble Space Telescope.



\section{Confirmation/rejection of candidates from photometric transit surveys}

Photometric observations help resolve tricky cases among candidates identified by photometric transit surveys, (a) by deciding whether a marginal transit detection is really a periodic transit-like signal, (b)  by getting a more precise shape to discriminate planetary transits from eclipsing binaries, (c) by identifying which of several stars included in the psf of the initial survey is undergoing the transit signal.

The challenge here is to measure the transit candidate at the correct epoch, with a bigger telescope than that used in the initial transit search, in order to increase the accuracy of the lightcurve. An extreme accuracy is usually not needed. Since the candidate was detected already in ``survey mode'' with a smaller aperture and lower sampling, a much better lightcurve is bound to result from measurements with a larger telescope.

\begin{figure}[!th]
\centering
\includegraphics[,width=5cm]{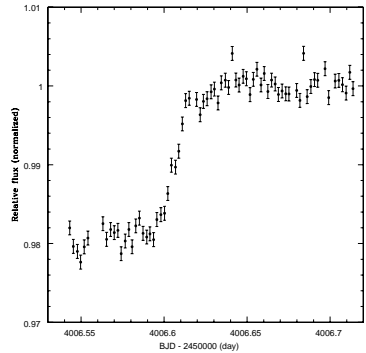}
\caption{The second half of the transit of WASP-2 measured from the 60-cm telescope at the St-Luc observatory.}
\label{stluc}
\end{figure}

 Examples of this mode abound from several transit surveys. The transits of TrES-1b where first confirmed with the IAC 80cm telescope before being measured in radial velocity. In the HATNet survey, based on small cameras, a dedicated larger telescope has been built, ``TopHAT'', in order to re-measure the transit candidates. The XO project relies on a group of amateur astronomer for the same purpose, while the tourism-oriented St-Luc 60cm in the Swiss Alps has been used to confirm the shape and phasing of the WASP-2 planet (Fig.~\ref{stluc}).

As illustrated by the diversity of participants and facilities, the motto in this topic could be ``it does not matter who you are, you just need to be at the right time at the right place''. The difficulty does not reside  in the photometry itself, but in the time constraint, combined with bad weather, day-night cycles and seasonal star visibility. An exciting consequence is that this allows low-cost facilities to make an important contribution to one of the frontline fields of astronomy.

\section{High-accuracy transit lightcurves for known transiting planets}

Once a transiting planet is detected, one of the first objectives is to
measure the radius of the planet to the level of a few percent, so that it can
constrain planetary structure and evolution models. Small variations in transit timing are also interesting as a way to detect further planets in the systems.

On paper, it is relatively easy to collect enough photons during the transit to measure the planet radius to a few percent and transit timing to a few seconds from transit lightcruves. In reality, however, this turns out to be difficult to achieve from the ground because of photometric systematics.

Beginning with the STIS/HST lightcurve for HD209458b by Brown et al. (2001), and from the ground with the FORS/VLT lightcurve for OGLE-TR-132b by Moutou et al. (2004), sub-millimagnitude lightcurves have now been collected for many transiting planets, and the capabilities and challenges of this mode are getting clearer.


\subsection{Transit depth and radius ratios}

 The star/planet radius ratio is proportional to the transit depth, and in principle 
can be measured precisely even with a transit lightcurve of moderate accuracy. In practice, however, it turns out to be unexpectedly tricky. Systematics affecting a transit lightcurve as a whole can cause significant errors in the determination of the transit depth without leaving any obvious trace in the lightcurve. This problem can be caused by the usual cast of atmosphere- and detector-related systematics in millimag ground-based CCD photometry. It can also be due to the difficulty of normalising lightcurves obtained by differential image analysis methods based on Alard (2000). 


\begin{figure}[!th]
\centering
\includegraphics[width=8cm]{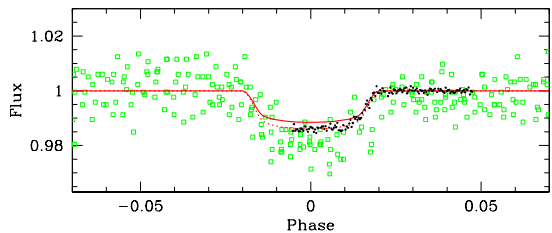}
\caption{Three sets of transit photometry for OGLE-TR-10 folded at the orbital period. The open symbols show the original OGLE photometry covering 7 individual transits (Udalski et al. 2002), the plain line shows the best-fit to the Magellan data of Holman et al. (2005), the squares and dotted line show the data and best-fit to VLT data from Pont et al. (2006).}
\label{ogle10}
\end{figure}

Figure~\ref{ogle10} illlustrates the  representative case of OGLE-TR-10, for which three sources give mutually incompatible transit depths The mismatches are much larger than the random errors and are all due to systematics in the photometry related to atmospheric changes (the effect of using different colour filters is negligible on this scale).


Another example is the discovery lightcurve for HD189733b (Bouchy et al. 2005) from the OHP 1.2m, which was shown by later measurements (Bakos et al. 2006) to be biased towards a higher depth by almost 30\%. In that case the cause turned out to be the paucity of comparison stars in the OHP images compounded by the airmass change.  


\subsection{The primary radius vs. orbital angle degeneracy}

There is a strong degeneracy in transit lightcurves between the size of the star and the angle of the orbit. Specifically, the product $R_\star M_\star^{-1/3}$ and the orbital angle can compensate each other to first order to produce a transit lightcurve of the same duration and similar shape.  

\begin{figure}[!th]
\centering
\includegraphics[width=9.5cm]{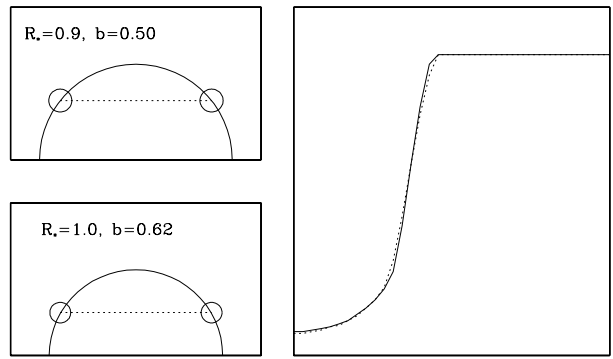}
\caption{The quasi-degeneracy between primary radius and orbital angle: sketch of the planet's path across the star (left) and transit lightcurve (right) for two configurations differing by 10\% on the star and planet size, with the orbital angle left free to compensate.}
\label{radiusangle}
\end{figure}

The orbital angle can be disentangled from $R_\star M_\star^{-1/3}$ with the exact shape of the transit ingress and egress, and the different behaviour of limb darkening in different colours. However, this requires photometry of exquisite precision. The difference between theoretical transit curves with different values of $R_\star$, when the other parameters are left to ajust, is generally smaller than 1~mmag, and significant only in very small portions of the phase (see Fig.~\ref{radiusangle}). Systematics in ground-based photometry can easily distort the signal.

\begin{figure}[!th]
\centering
\includegraphics[width=7cm]{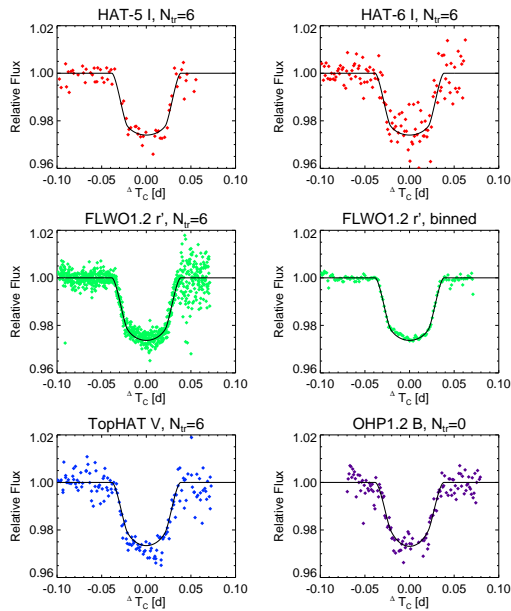}
\caption{Multi-site photometry for HD189733$b$ (reproduced from Bakos et al. 2006)}
\label{bakos}
\end{figure}

A concrete example was recently provided by the multi-site photometry for HD189733 presented by Bakos et al. (2006). This data, presented in Figure~\ref{bakos}, included a spectacular lightcurve from FLWO (upper left panel) that, using the formal errors, is able to define the shape of ingress and egress precisely enough to lift the degeneracy, and constrain the primary radius to 0.68 $\pm 0.02 R_\odot$. However, in the case of HD189733, we know that the stellar size is $R_\star =0.76 \pm 0.02 R_\odot$ (from the Hipparcos parallax and HST lightcurve). A posteriori, we can therefore estimate that the systematics introduced errors at least three times larger than the formal uncertainties.

\subsection{Transit timing variations}

The presence of a second planet in a transiting system can change the transit timings by up to several minutes. The monitoring of many individual transits is required to detect such a signal. Recent attempts include Steffen \& Agol (2005) for TrES-1,  Bakos et al. (2006) for HD189733, and Gillon et al. (2006) for OGLE-TR-113. An ambitious programme entitled "Transit Lightcurves Project" is attempting to monitor all transiting planets for timing variations (see Holman et al. in this volume). 

In all these studies, variations significantly larger than the random noise were found. However, they can also be attributed to atmospheric systematics. 


\subsection{Reduction methods}

To sum up, getting sub-millimagnitude lightcurves for transiting planets to lift the radius-angle degeneracy and study transit timing variations requires sub-millimagnitude accuracy both on the photon noise and on systematics. As sufficient photon noise is relatively easy to attain even on faint targets, the observing strategy and reduction must aim at reaching the lowest possible level of systematic noise. This leads us to the issue of the methods of photometric reductions.

There are four main reduction methods to obtain lightcurves accurate at the millimagnitude level for single objects with a CCD camera: (a) aperture photometry (b) PSF-fitting (c) differential image analysis (d) deconvolution. 

Aperture photometry is very fast, conceptually simple, and accurate as long as the field is not too crowded. It can yield reliable lightcurves even in moderately crowded fields if the results are then decorrelated for atmospheric variations. Unless there is a contaminant object in the close vicinity of the target, aperture photometry usually gives results that are only slightly more scattered than other more sophisticated methods, with a vast gain in computing time and simplicity.

Point-spread-function (PSF) fitting is tricky and shows a strong dependence on the quality of the PSF model and the actual shape and shape variation of the PSF on the images. It introduces a delicate subjective element in the choice of PDF stars and PSF parameters. As it represents a half-way compromise between aperture photometry and the methods more specifically adapted to differential lightcurves, it is generally not used in transit photometry.

Differential image analysis (DIA) has been the method of choice in recent years, using variations of the initial versions proposed by Alard \& Lupton (1999) and Alard (2000). It is specifically designed to detect small variations in a dense field, and therefore well-suited to transit lightcurves. It works by convolving a reference template image to the actual seeing of each image, scaling them, then subtracting them. In principle, this procedure subtracts out all the constant sources in the field and leaves only the variables.  

Deconvolution, recently adapted by Gillon et al. (2006) to transit lightcurves from an algorithm by Magain et al. (1998), works by deconvolving each image to a better-resolution image - such as would be obtained under much better seeing conditions (see Gillon et al. in this volume). It uses all the information in the field to reconstruct the PSF, and is conceptually sound. It can give spectacular results in very crowded fields. Its drawback is that it is at present exceedingly slow - requiring several days of CPU for the reduction of a few hours of data.


When comparing these four methods in the context of sub-millimagnitude transit lightcurves, the quality criterium is not, for the reasons explained above, which method produces the lightcurve with the smallest dispersion, but which one is most robust to correlated noise/systematics.

The choice of reduction method will be a case-by-case decision, and cross-checking the systematics with different methods applied to the same data often proves useful.

\subsection{Space telescope lightcurves}

Very robust photometry with systematics well below the $10^{-3}$ level can be obtained from space. Lightcurves from low-dispersion spectra were measured for HD209458 (Brown 2001) and TrES1 (R. Gilliland, priv. comm.). They are precise enough to disentangle the radius-angle degeneracy to the percent level, leaving only the $R M^{-1/3}$ ambiguity as the limiting factor on the planet size and density. 

\begin{figure}[!th]
\centering
\includegraphics[width=9.5cm]{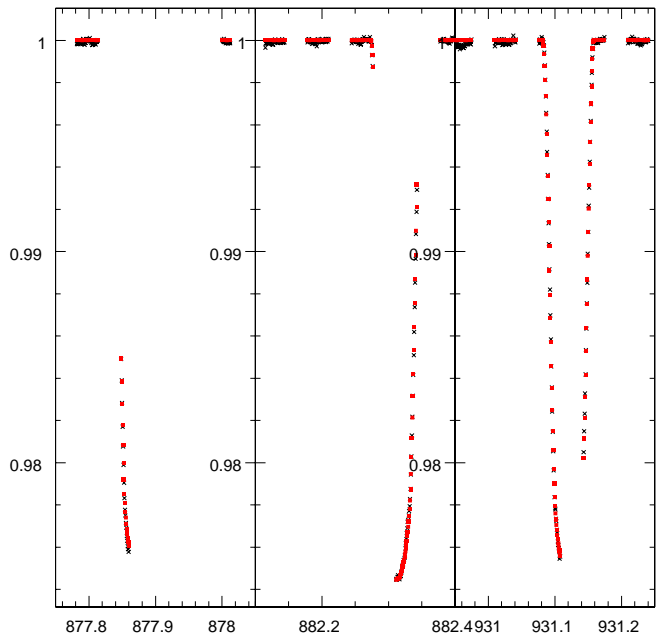}
\caption{HST lightcurve for HD189733 (crosses) with best-fit transit curve (squares). The dispersion of the residuals is 6$\times 10^{-5}$.}
\label{hst}
\end{figure}

We have recently obtained an HST lightcurve for HD189733, with residuals of only 60 {\it micro}mag on individual points (Fig.~\ref{hst}). This results in a planet radius determination with errors lower than 2\% - much less than the difference between the equatorial and polar radius of Jupiter. At this level of accuracy, several other features start becoming detectable: associated transit by Earth-sized moons, Saturn-type rings, extended planetary atmosphere, as well as granularity and spots on the host star, or  deviation from circularity of the star or planet.

\section{Conclusion and prospects}

{\it RV planets} \smallskip

As the radial-velocity surveys detect lighter planets in increasing numbers, there is a need for sub-millimagnitude photometry on 6-10 mag stars to detect transiting ice/rock planets. Ground-based photometry for bright stars is limited near the mmag level by the atmosphere-related systematics. The MOST satellite can contribute for the brightest targets, and the HST in a few well-constrained cases, but there is clearly a niche for a small space telescope dedicated to precision photometry on single objects.

\smallskip \noindent {\it Confimation/rejection of transit candidates} \smallskip

This is a well-established and relatively easy mode of observation, where the main limitation is telescope and operator time availability. This topic has contributed to giving a new role for small telescopes in front-line planet research.

This mode will be an essential component of the ground follow-up of the space-borne transit searches Corot and Kepler. Because the PSF of these missions covers many arcseconds, ground-based photometry will be needed to identify the background eclisping binaries,  expected to be the main source of contamination for planetary transit detection.

\newpage
\smallskip \noindent {{\it Parameters for known planets} \smallskip

Sub-millimagnitude stability is required, and rapid progress has been achieved in the past couple of years to obtain extremely precise lightcurves over a few hours with reduced systematics.  Impressive results have been obtained for both faint targets on 8-m class telescopes and bright targets with $<1$ m telescopes, with systematics smaller than 0.4 mmag. Still, unrecognised systematics can easily scatter planet size and transit timing measurements several formal sigmas away from the true value, making space measurements necessary. 

MOST and HST observations are planned for all presently known transiting planets brighter than V=12 mag. A dedicated space mission may be justified in the near future as ever lighter planets are found by radial-velocity surveys.

\acknowledgements 

We wish to thank Gaspar Bakos, Michael Gillon, Matthew Holman and Joel Hartman  for very insightful discussions related to the topic of this paper.


\end{document}